\documentclass[aps,prl,twocolumn,groupedaddress,amsmath,amssymb]{revtex4-1}
\usepackage{graphicx}
\usepackage{dcolumn}
\usepackage{bm}
\usepackage{epstopdf}
\usepackage{psfrag}
\usepackage{subfigure}
\usepackage{cancel}
\usepackage{color}

\begin{document}

\title{Repulsively induced photon super-bunching in driven resonator arrays}

\date{\today}

\author{T Grujic$^1$, S R Clark$^{2,1}$, D Jaksch$^{1,2}$ and D G Angelakis$^{2,3}$}
\address{$^1$ Clarendon Laboratory, University of Oxford, Parks Road, Oxford OX1 3PU, UK}
\address{$^2$ Centre for Quantum Technologies, National University of Singapore, 2 Science Drive 3, Singapore 117542}
\address{$^3$ Science Department, Technical University of Crete, Chania, Crete, Greece, 73100}
\email{t.grujic1@physics.ox.ac.uk, dimitris.angelakis@gmail.org}
\begin{abstract}
{\it We analyze the non-equilibrium behaviour of driven  nonlinear photonic resonator arrays under the selective excitation of specific photonic many-body modes. Targeting the unit-filled ground state, we find a counter-intuitive `super bunching' in the emitted photon statistics in spite of  relatively strong onsite repulsive interaction. We consider resonator arrays with Kerr nonlinearities described by the Bose Hubbard model, but also show that an analogous effect is observable in near-future experiments coupling resonators to two-level systems as described by the Jaynes Cummings Hubbard Hamiltonian.  For the  experimentally accessible case of a pair of coupled resonators forming a photonic molecule, we provide an analytical explanation for the nature of the effect.}
\end{abstract}

\maketitle
\emph{Introduction - }
Coupled resonator arrays (CRAs) offer the intriguing possibility of realising strongly-correlated many-body quantum states of light. Early work on CRAs assumed idealised, lossless arrays, and focussed in particular on equilibrium quantum phase transitions in these structures. However, near-future photonic devices will necessarily operate under driven-dissipative conditions on account of unavoidable photon loss, thereby serving as natural platforms for the exploration of novel non-equilibrium many-body photonic effects \cite{carusotto2009fermionized,angelakis2009steady,tomadin2010signatures,schmidt2010nonequilibrium,liew2010single,hartmann2010polariton,liew2010single,bamba2011origin,ferretti2010photon,knap2011emission}. Our understanding of these systems is in its infancy, making it desirable to concretely connect the non-equilibrium properties of CRAs with their more familiar equilibrium structure. To this end, there have been recent efforts to identify signatures of the  equilibrium quantum phase transition as originally proposed in \cite{angelakis2007photon,greentree2006quantum,rossini2007mott,hartmann2006strongly,aichhorn2008quantum,schmidt2009strong,koch2009superfluid,schmidt2010excitations,mering2009analytic} which survive under lossy dynamics. 

We propose an alternative scheme to chart different regions of parameter space and connect non-equilibrium observables to the underlying Hamiltonian properties. We envisage a resonator array driven to a non-equilibrium steady state (NESS) by external lasers, with the laser frequency chosen such that the unit-filled equilibrium ground state with on average one particle per site is selectively addressed and populated. Features arising from the details of the non-equilibrium operation appear in collected emission statistics, including a counter-intuitive many-body repulsion-induced bunching of the emitted photons, the magnitude of which is controllable via tuning Hamiltonian parameters. Novel super bunched light sources far exceeding the bunching of thermal photons may find important applications in ghost imaging technologies \cite{kolobov2007quantum} and all-optical simulation of two-photon correlations in quantum walks \cite{peruzzo2010quantum}. 

\begin{figure}[h]
  \centering
\includegraphics[width=0.49\textwidth]{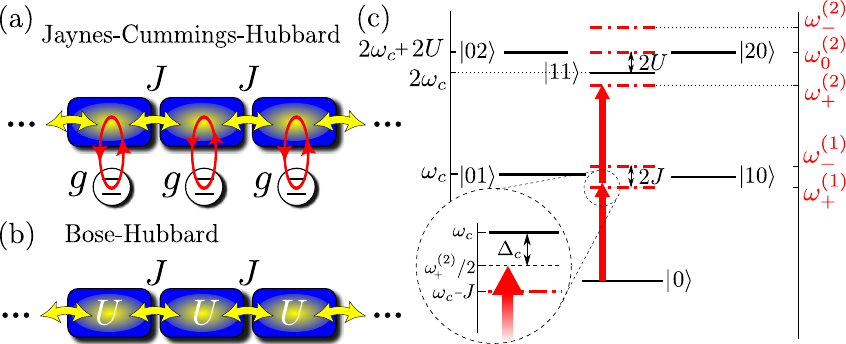}
\caption{(a) Schematic of the driven-dissipative Bose Hubbard model, featuring local coherent driving, photon tunnelling and a purely photonic Kerr nonlinearity. (b) The driven-dissipative Jaynes Cummings model with effective photon nonlinearity generated by couplings to two-level systems. (c) Diagram showing the bare basis (solid black) and eigen-frequencies (dash dotted red) of the driven system  for the minimal system of $M=2$ resonators. The laser is tuned so that two laser photons (vertical red arrows) are capable of promoting the system to the lowest-lying 2-photon state. }
\label{fig:sys_schematic}
\end{figure}

In this work we first focus on a minimal-sized two-site resonator system. Such dimers or a photonic molecules are expected to be experimentally viable in the near future in different technologies ranging from semiconductors to Circuit QED. We initially study the system for resonator nonlinearities of the repulsive Kerr-type, as shown in Fig.~\ref{fig:sys_schematic}~(a), to illustrate our driving scheme. We then analyze the Jaynes-Cumming  type encountered when resonator modes interact with embedded effective two-level systems \cite{fink2008climbing}, as in Fig.~\ref{fig:sys_schematic}~(b). We note here that we have previously investigated the validity of modelling the JCH with the simpler, single species Bose-Hubbard (BH) model \cite{grujic2012non}, and accordingly also demonstrate that the super-bunching signature persists under a JCH description. Moving beyond this minimal `array', bunched emission is also demonstrated in near future experimentally accessible mesoscopic-sized systems \cite {houck2012chip,majumdar2012design}. 

\emph{System - }
We consider a one-dimensional chain of $M$ coupled single-mode optical resonators under periodic boundary conditions. Each resonator of frequency $\omega_c$ is coherently coupled to its two nearest-neighbors. External lasers coherently drive each resonator in-phase with amplitude $\Omega$. In a frame rotating at the laser frequency $\omega_L$, the system Hamiltonian is:

\begin{eqnarray}
\hat{\mathcal{H}}^{(M)} & = & \sum_{j=1}^M \left [ \Delta_c \hat{a}_j^\dag \hat{a}_j + U \hat{a}_j^\dag \hat{a}_j^\dag \hat{a}_j \hat{a}_j + \Omega (\hat{a}^\dag_j + \hat{a}_j )\right ] \nonumber \\
& - & J \sum_{\langle j, j'\rangle} \left ( \hat{a}_j^\dag \hat{a}_{j+1} + \hat{a}_{j+1}^\dag \hat{a}_{j} \right ).
\label{eq:ham}
\end{eqnarray}
Here $U$ is the Kerr nonlinear strength, $J$ is the photon hopping rate, $\Omega$ is the photon driving strength, and $\langle j, j' \rangle$ denotes nearest neighbour resonators. The operators $\hat{a}_j$ are the photon destruction operators for the photon mode in resonator $j$. The detuning of the driving laser frequency from the bare cavity frequency is $\Delta_c = \omega_c - \omega_L$. 

Markovian photon loss processes from each cavity are incorporated via a quantum master equation formalism for the evolution of the system density matrix $\rho$, $\dot{\rho} = \mathcal{L}^{(M)} [\rho]$, where
\begin{equation}
\mathcal{L}^{(M)}[\rho] = \frac{1}{i} [\hat{\mathcal{H}}^{(M)}, \rho] + \gamma_p \sum_{j=1}^M \hat{\mathcal{D}}_{\hat{a}_j}[\rho].
\label{eq:master_equation}
\end{equation}
The dissipative part of the dynamics is described by $\hat{\mathcal{D}}_{\hat{O}}[\rho] = \hat{O} \rho \hat{O}^\dag - \frac{1}{2} (\hat{O}^\dag \hat{O} \rho + \rho \hat{O}^\dag \hat{O} )$. The NESS of the system is described by the density matrix $\rho_{\rm ss}$ which satisfies $\mathcal{L}[\rho_{\rm ss}] = 0$, and observables are measured with respect to this state, $\langle \hat{O} \rangle_{\rm ss} \equiv {\rm Tr} (\hat{O} \rho_{\rm ss})$. 

\emph{Two-resonator `dimer' - }
We begin by analyzing the simplest possible driven resonator `array' consisting of just $M=2$ resonators, which serves to illustrate clearly our scheme for accessing the unit-filled ground state. Figure~\ref{fig:sys_schematic}~(b) shows the low-lying eigenstructure of the undriven Hamiltonian of Eq.~(\ref{eq:ham}) for $M=2$, and our driving scheme. The two one-photon eigen-frequencies are the symmetric (+) and anti-symmetric (-) Bloch modes, and lie (in the bare frame, with $\Omega = 0$) at $\omega^{(1)}_\pm = \omega_c \mp J$, with corresponding eigenstates $|1_\pm\rangle$. The two-photon eigen-frequencies are
$\omega_0^{(2)} = 2 \omega_c + 2U$, 
$\omega_\pm^{(2)} = 2 \omega_c + U \mp \sqrt{U^2 + 4 J^2}$, with eigenstates $|2_0\rangle$, $|2_\pm \rangle$, respectively. The unit-filled ground state is of frequency $\omega_+^{(2)}$. This mode undergoes a qualitative change between the extreme limits of a localized state, characterized by vanishing on-site photon number fluctutation ${\rm Var} (\hat{a}_j^\dag \hat{a}_j) \rightarrow 0$ for $U \gg J$, to a coherent superposition state with ${\rm Var} (\hat{a}_j^\dag \hat{a}_j) \rightarrow \frac{1}{2}$ for $U \ll J $. For increasing system size, the behavioural transition of the ground state becomes sharper, approaching the celebrated Bose Hubbard Mott-insulating to superfluid phase transition in the infinite system limit \cite{jaksch1998cold}. 

To selectively populate the unit-filled ground state of Eq.~(\ref{eq:ham}), we set the driving laser frequency such that two laser photons are resonant with the lowest-lying two photon mode, i.e. $2 \omega_L = \omega_+^{(2)}$, implying a laser detuning
\begin{equation}
\Delta_c (J, U) = \frac{1}{2} \left ( \sqrt{U^2 + 4 J^2} - U \right ).
\label{eq:laser_detuning}
\end{equation}
Fixing $\Delta_c = \Delta_c (J,U)$ for a given hopping and nonlinearity, we now examine which features of the underlying Hamiltonian mode structure leave fingerprints on experimentally accessible photonics observables. In our non-equilibrium setting the photon number is not an integer and the photon number variance is not an informative order parameter. To compensate for these non-equilibrium effects, we instead focus on the local zero-time photon correlation function $g^{(2)} \equiv g^{(2)}_j = \langle \hat{a}_j^\dag \hat{a}_j^\dag \hat{a}_j \hat{a}_j \rangle / \langle \hat{a}_j^\dag \hat{a}_j \rangle^2$, a statistical quantity directly accessible in CRA setups through standard methods like homodyne detection. We note that $g^{(2)}$ measurements may be particularly valuable in weakly-driven systems, where the excitation number may be very small, but normalised statistical quantities may be collected over longer times. 

\begin{figure}[t]
  \centering
\subfigure{\includegraphics[width=0.48\textwidth]{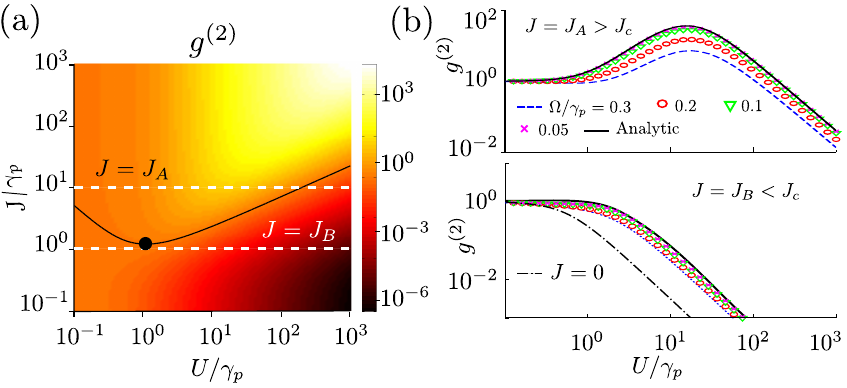}}
\caption{(a) Emitted photon statistics as a function of $J$ and $U$  for a dimer ($M=2$) of resonators driven according to Eq.~(\ref{eq:laser_detuning}). Black curve: analytic result for $g^{(2)}=1$. Black dot: the point $(J_c/\gamma_p, U_c/\gamma_p)$ where bunching sets in. (b) `Slices' along the dotted lines in (a), above and below the critical hopping amplitude at which the bunching feature appears, for decreasing driving strengths, down to the infinitesimal limit (solid black lines).} 
\label{fig:phase_diagrams}
\end{figure}

Figure \ref{fig:phase_diagrams}~(a) shows $g^{(2)}$ measured in the NESS of a BH dimer pumped at the laser detuning of Eq.~(\ref{eq:laser_detuning}), for a range of tunnelling rates and nonlinearities $(J,U)$. The diagram is broadly  divided into three regions, defined by Poissonian ($g^{(2)} \approx 1$), anti-bunched ($g^{(2)} < 1$) and bunched ($g^{(2)} > 1$) statistics. Notably, there is a critical coupling rate $J_{\rm crit}$ between the resonators below which bunching does not occur for any value of nonlinearity, suggesting that the bunching arises from a co-operative many body effect in the NESS. Figure \ref{fig:phase_diagrams}~(b) shows the qualitative difference in the behaviour of the correlation function above and below this critical point. For comparison, an isolated resonator driven at it's single-particle (unit-filled) resonance never exhibits bunched signatures (see dash-dotted line in lower panel of Fig.~\ref{fig:phase_diagrams}~(b)).

At low nonlinearities $U \ll \gamma_p$, the dimer is driven at the frequency of the zero-momentum Bloch mode, and its response is approximately linear. The NESS is a coherent state, inheriting Poissonian statistics with $g^{(2)} = 1$, and average population $\langle \hat{a}_j^\dag \hat{a}_j \rangle = (2 \Omega / \gamma_p)^2$, for all $J$. At the other extreme, taking the hardcore limit $U \rightarrow \infty$, no more than a single photon per resonator can be injected, and the problem is reduced to two coupled two-level systems whose emission is completely anti bunched ($g^{(2)} = 0$), with mean excitation number per resonator
$\lim_{U \rightarrow \infty} \langle \hat{a}^\dag \hat{a} \rangle = x (2 x + 1)/((2 x + 1)^2 + x y)$, where $x \equiv (2 \Omega / \gamma_p )^2$ and $y \equiv (J / \Omega)^2$. 

Away from these extreme limits, in the region of parameter space where $J$ and $U$ are comparable, the emitted light is bunched, for hopping rates larger than a critical rate $J > J_c$. This is counter-intuitive, as we are probing a two-photon state with significant repulsion favouring separation, and yet we find an enhanced probability of photons being emitted together, relative to the statistics of the driving. 

We derive features of this bunched region by considering the limit of infinitesimal driving. Figure~\ref{fig:phase_diagrams}~(b) shows that the correlation function approaches a limiting behaviour as $\Omega / \gamma_p \rightarrow 0$, an observation confirmed by perturbatively expanding the elements of the NESS density matrix $\rho_{\rm ss}$ in increasing powers of the driving strength, then solving for $g^{(2)}$ \cite{ferretti2010photon}. The equations thus obtained are physically opaque, however further progress can be made by making a pure-state ansatz $\rho_{\rm ss} = |\Psi_{\rm ss} \rangle \langle \Psi_{\rm ss} |$, valid in the low driving regime \cite{bamba2011origin}. The state $| \Psi_{\rm ss} \rangle$ is found as the stationary state (corresponding to the zero eigenvalue) of the effective Hamiltonian $\hat{\mathcal{H}}^{(M)}_{\rm eff}$ obtained by replacing $\Delta_c \rightarrow \Delta_c - i \gamma_p /2$ in Eq.~(\ref{eq:ham}), which may be interpreted as the Hamiltonian governing a single quantum trajectory, with a vanishing probability of a quantum jump ensured by taking the limit $\Omega / \gamma_p \rightarrow 0$. 

Considering only the lowest excitations and exploiting the symmetry of the two-site system, we set $| \Psi_{\rm ss} \rangle = C_{00} |00\rangle + C_1(|01\rangle + |10 \rangle) + C_{11} |1 1\rangle + C_2(|02\rangle + |20 \rangle)$, from which the correlation function follows as $g^{(2)} = |C_2|^2 / |C_1|^4$. We obtain the minimum resonator coupling for which bunched statistics appear as $\tilde{J}_c = J_c / \gamma_p = \sqrt{(3 + 2 \sqrt{2})/4}$, and the nonlinearity at this point $\tilde{U}_c = U_c / \gamma_p = \sqrt{\tilde{J}_c}$ as marked in Fig.~\ref{fig:phase_diagrams}~(a) (see supplementary material for further details). The transition from super- to sub-Poissonian statistics (i.e. where $g^{(2)}=1$) is found to occur at $\tilde{J} \approx \sqrt{\tilde{U} / 2}$ for $J \gg J_c$, while for very large but finite $U / J$, the exact solution may be simplified to $g^{(2)} \approx \left ( \frac{\tilde{J}}{\tilde{U}} \right )^2 \left ( 1 + 4 \tilde{J}^2 \right )$.

\begin{figure}[h]
  \centering
\includegraphics[width=0.49\textwidth]{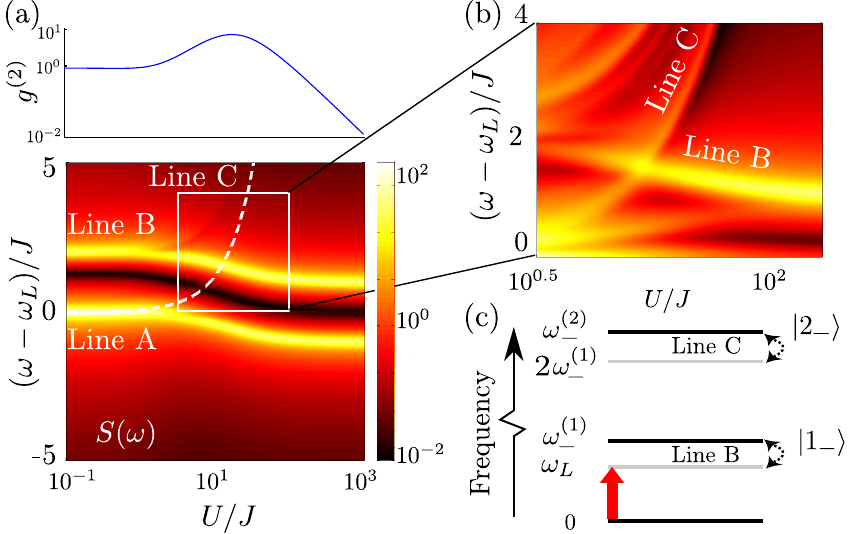}
\caption{(a) Lower: Emission spectral function relative to the laser frequency $|S(\omega - \omega_L)|^2$ with increasing nonlinearity $U$ at a fixed hopping $J / \gamma_p = 10$. At our weak driving ($\Omega / \gamma_p = 0.3$), the spectrum is dominated by transitions between the ground and one-particle manifold (bright lines A and B). Weaker features involving the two-particle manifold are also present, for instance the dashed curve highlighting line C. Upper: the zero-time correlation function $g^{(2)}$. (b) Magnification of the region inside the white square in (a), calculated at the higher driving $\Omega / \gamma_p = 1$ to highlight the crossing of the features labelled Lines B and C. (c) Transitions involved in spectral lines B and C, with only the relevant modes drawn. }
\label{fig:explanation}
\end{figure}

Figure~\ref{fig:explanation} offers physical insight into this phenomenon, showing the emission spectrum of the system as calculated from the Fourier transform $S(\omega - \omega_L)$ of the on-site steady-state auto-correlation function $S(\tau) = \langle \hat{a}^\dag (t + \tau) \hat{a} (t) \rangle$, as a function of increasing nonlinearity. The resonator coupling is sufficiently large ($J / \gamma_p = 10 > \tilde{J}_c$) to observe bunched emission (top panel of Fig.~\ref{fig:explanation}~(a)). At all nonlinearities, the spectrum is dominated by two bright features. These correspond to decays from the lower and upper one-particle states $|1_\mp \rangle$ to the vacuum $|0\rangle$, labelled lines A and B respectively. 

Weaker features are also present, which do not significantly affect the steady state photon populations, but may strongly modify statistical quantities such as $g^{(2)}$. Line C in Figs.~\ref{fig:explanation}~(a) and (b) corresponds to emission from the highest two-particle state $|2_+\rangle$ to an intermediate level, as drawn in Fig.~\ref{fig:explanation}~(c). In the vicinity of the crossing of emission lines B and C at $\tilde{U} / \tilde{J} = (9 - \sqrt{17})/4$, the population in $|2_-\rangle$ reaches a maximum, as photons emitted as part of the line B decay process may transfer population instead to $|2_-\rangle$. Further, calculating projections into the bare basis, we find that there is a greater probability of finding the two photons of the mode $|2_-\rangle$ in the same resonator than distributed between them, relative to the driven $|2_+\rangle$ mode. This is reflected in the enhanced probability of simultaneous emission of two photons ($g^{(2)}>1$) around this crossing. In contrast the mode $|2_+\rangle$ favours delocalising its two photons relative to the statistics of the driving laser. Thus, we observe either approximately coherent, or anti-bunched light in all regions of parameter space except in the vicinity of the crossing of lines B and C. For resonator couplings $J$ less than the line width $\gamma_p$ of the resonators, the global physics of the system resembles that of an isolated nonlinear resonator driven at it's resonance frequency, such that anti-bunching is always expected -- in spectral terms, the crossing of lines B and C is hidden inside the coalesced lines A and B. 

\emph{Larger systems - }
We now investigate how the correlations presented in Fig.~\ref{fig:phase_diagrams}~(b) evolve as the system size increases, continuing to drive the commensurately-filled ground state. An analytic approach valid for arbitrary Hamiltonian couplings beyond $M=2$ resonators is intractable. Instead, we numerically calculate the eigenvector of the effective Hamiltonian $\hat{\mathcal{H}}^{(M)}_{\rm eff}$ with eigenvalue closest to zero (taking a series of successively weaker drivings $\Omega / \gamma_p$ to ensure convergence of observables). Exploiting the translational invariance of systems with periodic boundary conditions allows us to access systems of up to $M=7$ resonators while retaining three photons per resonator in simulations. Rigorous quantum trajectory calculations based on the matrix product state representation and the time evolving block decimation algorithm \cite{vidal2003efficient,vidal2004efficient,white2004real} performed at a finite driving strength (see Supplementary material) broadly agree with the results obtained via numerically exact diagonalisation of $\hat{\mathcal{H}}^{(M)}_{\rm eff}$. However, they do indicate that the precise features of the bunching region do depend on drive strength, as already observed for $M=2$ in Fig.~\ref{fig:phase_diagrams}~(b). 

\begin{figure}[h]
  \centering
\includegraphics[width=0.48\textwidth]{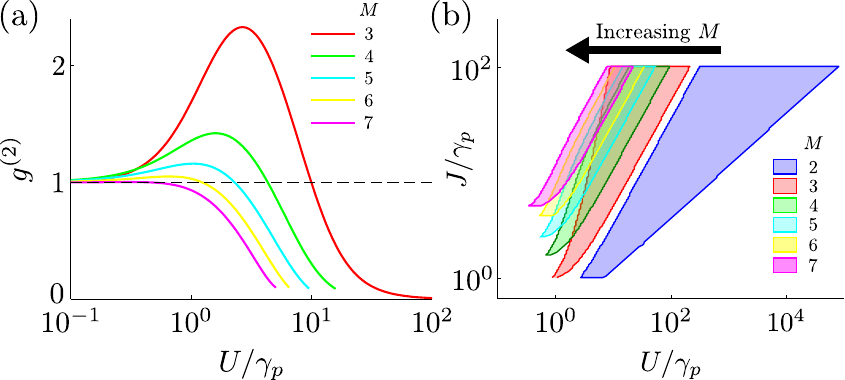}
\caption{ (a) Correlation function evaluated as a function of increasing nonlinearity at fixed resonator coupling $J / \gamma_p = 10^1$. Note the $M=2$  results are not included as the bunching in this case is significantly larger. (b) Extent of the bunched region in $(J,U)$ parameter space, as measured from $U=U_{\rm LHS}$ corresponding to the peak $g^{(2)}$, to $U=U_{\rm RHS}$ at which the correlations change from bunched to anti-bunched, for a range of resonator couplings $J$. }
\label{fig:increasing_system_size}
\end{figure}

Figure~\ref{fig:increasing_system_size} shows the evolution of the counterintutive bunched region for increasing system sizes up to $M=7$ resonators. We see a reduction in the magnitude and range of interaction strengths for which bunched light is observed as the system size increases. The bunching region is seen to retreat up the $J$-axis, while smaller interaction strengths $U$ are necessary to induce the bunching. This explains the reduction in the magnitude of the effect observed for cross-sections at constant resonator coupling, as in Figure~\ref{fig:increasing_system_size}~(a).

\emph{Photon statistics in Jaynes Cummings arrays - }
Near future circuit QED systems will most probably realise a few-photon resonator nonlinearity via a Jaynes-Cummings interaction with embedded two-level systems \cite{angelakis2007photon}, as described by the Jaynes Cummings Hubbard  Hamiltonian in a rotating frame
\begin{eqnarray}
\hat{\mathcal{H}}_{\rm JCH}^{(M)} & = & \sum_{j=1}^M \left ( \Delta_c \hat{a}_j^\dag \hat{a}_j + (\Delta_c - \Delta) \hat{\sigma}_j^+ \hat{\sigma}_j^-  + \Omega (\hat{a}^\dag_j + \hat{a}_j) \right )\nonumber \\
& - & J \sum_{<j,j'> } \left (\hat{a}_j^\dag \hat{a}_j' \right ).
\label{eq:ham_JCH}
\end{eqnarray}
Here $\Delta = \omega_c - \omega_a$ denotes the difference between the resonators' frequency and the TLS transition frequency, and $\hat{\sigma}^\pm$ denote TLS raising and lowering operators. The JCH is known to possess a localized-delocalized transition as either the hopping $J$ is increased, or the Jaynes-Cummings parameter $\Delta$ is made more negative. This transition is similar in some respects to the phase transition of the BH model, though also differs in fundamental ways on account of the different nature of the systems's intrinsic excitations (bosons and polaritons, respectively) \cite{aichhorn2008quantum,schmidt2009strong,koch2009superfluid,schmidt2010excitations,mering2009analytic,grujic2012non}. 

\begin{figure}[h]
  \centering
\includegraphics[width=0.48\textwidth]{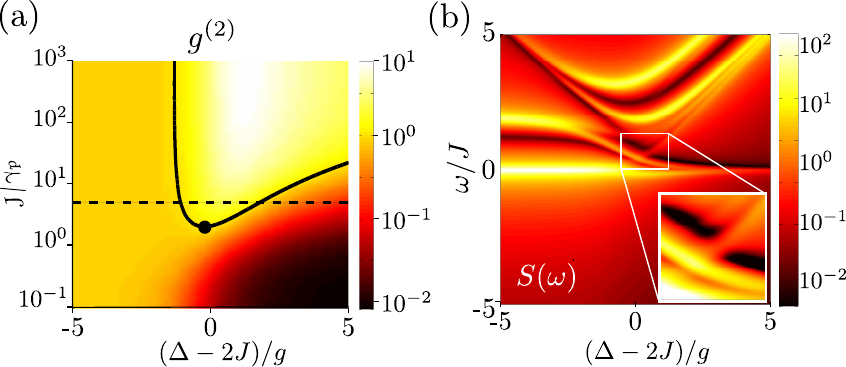}
\caption{Steady state observables for a two-site Jaynes-Cummings array, driven at its lowest two-particle resonance. Parameters: $g / \gamma_p = 10$. (a) Zero-time photon correlation as a function of resonator coupling $J$ and atom-resonator detuning $\Delta$, playing the part of an effective photon nonlinearity. (b) Spectral function $|S(\omega)|^2$ evaluated along the dashed line in (a), again showing a crossing of emission lines promoting photons to a state with enhanced probability of bunched emission (red arrow is the driving laser frequency). }
\label{fig:JCH}
\end{figure}
Figure~\ref{fig:JCH} presents evidence that the mechanism underlying the bunched emission discussed above in the context of a the driven Bose-Hubbard model persists in this qualitatively different setting for realistic atom-resonator couplings and loss rates, and is therefore observable in near future state of the art experiments involving just two coupled resonators.

\emph{Conclusions - }
We have proposed the selective excitation of photonic many-body modes of interest in open resonator arrays using external driving lasers, over which we have full control of frequency and amplitude. We have shown how a combination of the equilibrium Hamiltonian structure and non-equilibrium operation lead to an interaction-induced region of bunched emission. This feature was found to persist in mesoscopic-sized arrays, and is also found under a more realistic array description, making its observation feasible in coming experiments. 

\bibliography{bibtex_bibliography}

\end{document}